\documentclass[aps,prl,twocolumn,groupaddress]{revtex4}
\usepackage{graphicx}
\usepackage{dcolumn}
\usepackage{bm}
\usepackage{color}
\begin{document}
%
\title{Current-induced persistent magnetization in a relaxorlike manganite}
%
\author{H. Sakai$^1$ and Y. Tokura$^{1,2,3}$}
\affiliation{$^1$Department of Applied Physics, University of Tokyo, Tokyo 113-8656, Japan\\$^2$Cross-Correlation Materials Research Program (CMRG), RIKEN, Wako 351-0198, Japan\\$^3$Multiferroics Project, ERATO, Japan Science and Technology Agency (JST), Tokyo 113-8656, Japan}
%
\begin{abstract}
A single crystal of 7\% Fe-doped (La$_{0.7}$Pr$_{0.3}$)$_{0.65}$Ca$_{0.35}$MnO$_3$ shows up as a typical {\it relaxor ferromagnet}, where ferromagnetic metallic and charge-orbital-ordered insulating clusters coexist with controllable volume fraction by external stimuli.
There, the persistent ferromagnetic metallic state can be produced by an electric-current excitation as the filamentary region, the magnetization in which is increased by $\sim\!0.4$ $\mu_{\rm B}$ per Mn.
A clear distinction from the current heating effect in a magnetic field, which conversely leads to a decrease in ferromagnetic fraction, enables us to bi-directionally switch both the magnetization and resistance by applying the voltages with different magnitudes.
\end{abstract}
%
\maketitle
%
Correlated electron systems with competing ordered states often show a dramatic phase change in response to even a minute external stimulus.
One such example is a hole-doped perovskite manganite, in which a keen competition  between a charge-orbital-ordered insulating (CO) phase and a ferromagnetic metallic (FM) one is a major source for magnetic-field-induced phase transitions or colossal magnetoresistance (CMR) phenomena.\cite{Tokura2006review}
The difference in free energy between the two states is so small near the bicritical phase boundary that an external perturbation other than a magnetic field can also drive the insulator-metal transition.\cite{Asamitsu1997NatureA,Miyano1997PRLa,Kiryukhin1997NatureA,Moritomo1997PRBa}
Among them, the electric-field-induced collapse of the CO into the conducting, perhaps FM, state has been intensively investigated since its first discovery in Pr$_{0.7}$Ca$_{0.3}$MnO$_{3}$ crystal.\cite{Asamitsu1997NatureA}
For such a current excitation, there has always been some concern about the Joule heating since the induced low-resistance state is occasionally similar in a resistance value to the high-temperature state due to the inherent semiconducting nature.
In CMR manganites, the observation of {\it persistent} changes in magnetization as well as in resistance may be the direct evidence for the genuine electric action on the phase change, although such studies have been rare\cite{Stankiewicz2000PRBa,Guha2000PRBa,Garbarino2006PRBa}.
In particular, the current production of the persistent ferromagnetic state is also an intriguing issue in the light of electromagnet function in nanometric materials.
%
\par
%
In this Letter, we have investigated electric-current-induced effects on a CMR manganite, especially on Fe-doped (La$_{0.7}$Pr$_{0.3}$)$_{0.65}$Ca$_{0.35}$MnO$_3$.
The Fe doping suppresses the FM correlation while stabilizing the CO one, leading to the phase-separated ground state with the FM and CO clusters coexisting.\cite{Sakai2007PRBa}
The feature is analogous to the case of Cr-doped CO manganites\cite{Raveau1997JSSCa,Kimura1999PRLa}, although the charge-orbital order remains short-ranged ($\le\!2$ nm)\cite{Sakai2007PRBa} and forms the macroscopically near-isotropic lattice and electronic structure in the present case.
Such a system has been known as a {\it relaxor ferromagnet},\cite{Kimura1999PRLa} showing glassy magnetic and magnetotransport properties, such as magnetic-field annealing effects and long-time relaxation phenomena.
Furthermore, the controlled volume fraction of the FM and CO states can be durably fixed at low temperatures due to the multistability characteristic of the relaxor system.
This will give an ideal arena for investigating the intrinsic effect of external perturbations accompanying temporary heating, since the persistent change can be detected free from the heating problem, as was done in the photoirradiation case.\cite{Okimoto2002APLa}
We here present a direct evidence for the current-induced persistent FM state in distinction from the heating effect in the relaxor ferromagnet.
%
\par
\begin{figure}
\includegraphics[width=8.5cm]{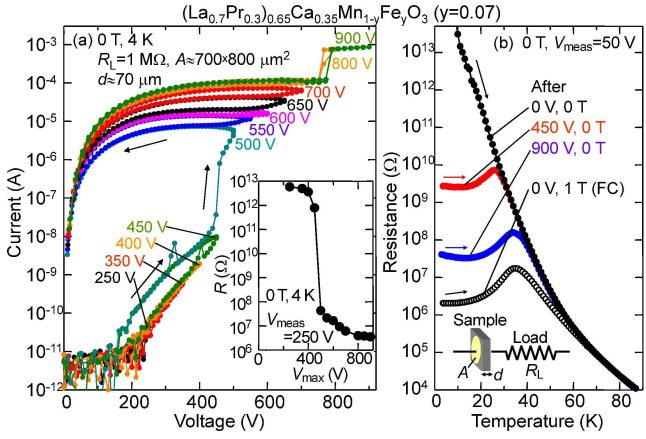}
\caption{\label{fig:0T}(Color online) (a) $I$-$V$ characteristics for a single crystal of (La$_{0.7}$Pr$_{0.3}$)$_{0.65}$Ca$_{0.35}$Mn$_{0.93}$Fe$_{0.07}$O$_3$ at 4 K in zero magnetic field. Inset: change in resistance $R$ as a function of the maximum voltage $V_{\rm max}$ (resistance-sensing voltage $V_{\rm meas}\!=\!250$ V). (b) Temperature profiles of resistance ($V_{\rm meas}\!=\!50$ V) in warming runs at 0 T after applying $V_{\rm max}$ (0, 450 and 900 V) at 4 K. Data measured after magnetic-field cooling (FC) at 1 T are also shown. Inset: a schematic diagram of the circuit; the electrode distance $d\!\sim\!70$ $\mu$m, the electrode area $A\!\sim700\times800$ $\mu$m$^{2}$, and the load resistance $R_{\rm L}\!=\!1$ M$\Omega$.}
\end{figure}

%
A single crystal of (La$_{0.7}$Pr$_{0.3}$)$_{0.65}$Ca$_{0.35}$Mn$_{0.93}$Fe$_{0.07}$O$_3$ was grown by the floating zone method, as detailed elsewhere\cite{Sakai2007PRBa}.
Electrodes, made of heat-treatment-type silver paint, were formed on the both end surfaces of the crystal specimen.
Current-voltage ($I$-$V$) characteristics were measured for the crystal connected in series with the load resistor ($R_{\rm L}\!=\!1$ M$\Omega$) [see the inset to Fig. \ref{fig:0T}(b)].
The sample resistance $R$ was measured at a constant voltage $V_{\rm meas}$ with a two-probe configuration; the contact resistance was $\sim$10 $\Omega$ and can be safely ignored in the low-temperature high-resistance state.
For the simultaneous measurement of the magnetization $M$ and $R$, the sample mounted in a superconducting quantum interference device magnetometer was connected to the $I$-$V$ measurement system through a co-axial cable.
%
\par
%
%
Figure \ref{fig:0T}(a) shows the typical $I$-$V$ characteristics at 4 K for a crystal of (La$_{0.7}$Pr$_{0.3}$)$_{0.65}$Ca$_{0.35}$Mn$_{0.93}$Fe$_{0.07}$O$_3$, which is insulating down to the lowest temperature [see Fig. \ref{fig:0T}(b)].
For this measurement, the sample was first cooled to 4 K at 0 T and then a bias voltage was swept as 0 V$\rightarrow V_{\rm max}\rightarrow 0$ V.
This sweep was repeated with successively increasing $V_{\rm max}$ from 250 V to 900 V.
For $V_{\rm max}\!\le\!450$ V, the system remains at the initial high-resistance state, exhibiting the nonlinear $I$-$V$ curve but with no (or minimal) hysteresis.
For $V_{\rm max}\!=\!500$ V, however, the $I$-$V$ curve displays a marked hysteresis, indicating that the system is switched to the low-resistance state with the $R$ drop by $\sim$5 orders of magnitude.
Similar hysteretic curves with smaller widths are also observed when $V_{\rm max}$ is further increased up to 700 V, accompanying a gradual decrease in $R$ by another one order of magnitude.
For $V_{\rm max}\!\ge\!800$ V, a sudden increase of current occurs at $\sim$750 V probably due to the current heating effect.
In fact, this lower-resistance state is not persistent and easily vanishes for $V\!\le\!750$ V.
In the high-voltage regime, the current heating would transiently raise the local temperature of the crystal above the glass transition temperature ({\it vide infra}), leading to no further persistent decrease in $R$.
The inset to Fig. \ref{fig:0T}(a) summarizes the change in $R$ versus $V_{\rm max}$, measured after each voltage sweep.
%
\par
%
We show in Fig. \ref{fig:0T}(b) the temperature profiles of $R$ in warming runs at 0 T after applying the bias voltage (0, 450, and 900 V) at 4 K at 0 T.
As a comparison, we also display the corresponding data measured after magnetic-field cooling (FC) at 1 T.
Regardless of a current or a magnetic field, the induced low-resistance state survives only at low temperatures (below $\sim$50 K), where the system exhibits a spin-glass phase, and it returns to the pristine high-resistance state by warming temperature.
Therefore, the effect of a current excitation should be attributed not to the dielectric breakdown caused by some permanent lattice-structural damage, but to the modification of the local electronic/lattice state, i.e., the persistent increase in the metallic volume fraction.
%
\par

\begin{figure}
\includegraphics[width=7cm]{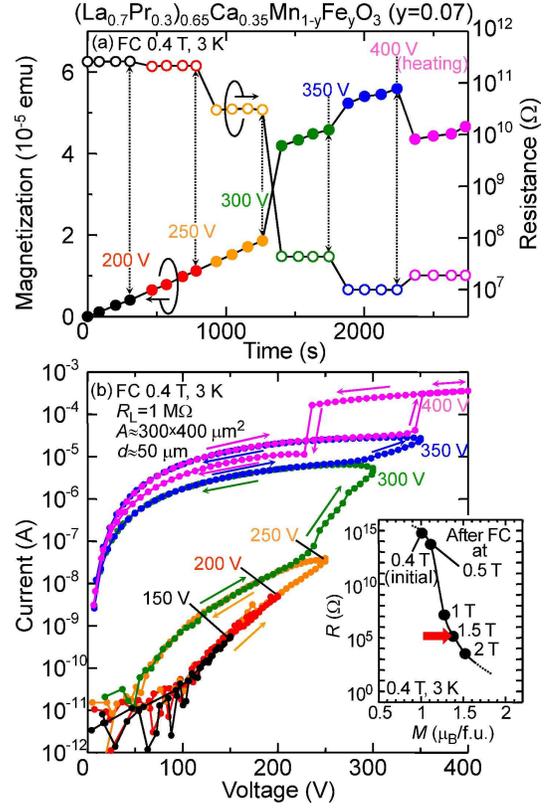}
\caption{\label{fig:04T}(Color online) (a) Variation of magnetization ($\Delta M$) and resistance ($V_{\rm meas}\!=\!150$ V) for a single crystal of (La$_{0.7}$Pr$_{0.3}$)$_{0.65}$Ca$_{0.35}$Mn$_{0.93}$Fe$_{0.07}$O$_3$ ($d\!\sim\!50$ $\mu$m, $A\!\sim\!300\!\times\!400$ $\mu$m$^{2}$) at 3 K at 0.4 T after FC. Dotted vertical arrows indicate the timing for applying bias voltages. (b) $I$-$V$ characteristics corresponding to the bias voltage sweeps ($R_{\rm L}\!=\!1$ M$\Omega$). Inset: resistance $R$ ($V_{\rm meas}\!=\!50$ V) versus magnetization $M$ at 3 K at 0.4 T after FC with various magnitudes of magnetic fields.}
\end{figure}
%
To directly detect the possible ferromagnetism in the current-induced metallic state, we have concurrently measured $M$ and $R$ with applying bias voltages.
Figure \ref{fig:04T}(a) displays their variation at 3 K after FC at 0.4 T.
Dotted vertical arrows and voltage values indicate the timing for the voltage sweep and its $V_{\rm max}$, respectively, where $V_{\rm max}$ was successively increased. 
The corresponding $I$-$V$ curves are shown in Fig. \ref{fig:04T}(b).
$M$ is barely saturated at 0.4 T while the system shows still insulating down to 3 K.
Note that temporal drift of $M$ reflects the slow dynamics in the glassy system.
When applying $V_{\rm max}\!\le\!250$ V, no change was discerned in $M$ while $R$ exhibits a small drop.
By further increasing $V_{\rm max}$ up to 300 and 350 V, we observed a marked increase in $M$ as accompanied by a large drop in $R$.
This ensures that the FM state is induced persistently by a current excitation, which we hereafter call the ``intrinsic" effect.
After applying $V_{\rm max}\!=\!400$ V, on the other hand, we found a decrease in $M$ as well as an increase in $R$.
Such a decrease in FM fraction may stem from the {\it rapid} temperature cycle, consisting of the abrupt temperature rise due to the Joule heating and the subsequent rapid cooling back to 3 K.
This was already observed for $V_{\rm max}\!=\!400$ V in Fig. \ref{fig:04T}(b) as a sharp increase in current at $\sim$350 V in a voltage-increasing run and a decrease at $\sim$230 V in a voltage-decreasing run.
The slowly field-cooled state should be the least affected in principle by the {\it slow} similar thermal cycles.
However, the rapid cooling may lead to a smaller volume fraction of the FM state, since its evolution induced by a magnetic field is subject to the long-time relaxation effect.\cite{Kimura1999PRLa}
Note that the Joule heating similarly observed at zero field (for $V_{\rm max}\!\ge\!800$ V) gives no influence on the FM fraction, as shown in Fig. \ref{fig:0T}(a).
Thus, the heating phenomenon in a magnetic field is clearly separated as the extrinsic electric-current effect from the intrinsic one; they have the opposite impacts on the FM volume fraction.
%
\par
%
The total increase in $M$ induced by the intrinsic current effect amounts to $2.0\!\times\!10^{-5}$ emu at 3 K at 0.4 T.
In case of the FM state produced homogeneously over the crystal, the net current-induced $M$ would be as small as $3.3\!\times\!10^{-3}$ $\mu_{\rm B}$ per Mn.
This is, however, completely contradictory to the large $R$ drop by $\sim\!5$ orders of magnitude.
It is likely that the FM state is induced only locally, forming the filamentary path between the electrodes.
When the FM clusters are evenly distributed over this crystal, the values of $M$ and $R$ at 3 K at 0.4 T should have the relation shown in the inset to Fig. \ref{fig:04T}(b), which was obtained by changing the FM fraction with varying the magnitude of a magnetic field in FC.
Provided that only the filamentary region formed by current flowing changes the FM fraction obeying the above relation, we can roughly estimate its cross-sectional size so as to make the experimental results self-consistent; this leads to the conclusion that the FM state with $M\!\sim\!1.4$ $\mu_{\rm B}$ per unit cell and $R\!\sim\!1.4\!\times\!10^{5}$ $\Omega$ [indicated by the thick horizontal arrow in the inset to Fig.\ref{fig:04T}(b)] should be formed in $\sim\!1/120$ of the area of the crystal ($\sim\!400\!\times\!300$ $\mu$m$^2$).
Consequently, $\Delta M$ induced in such a filamentary region is $\sim\!0.40$ $\mu_{\rm B}$ per Mn, being comparable to the photoinduced case.\cite{Okimoto2002APLa}
The estimated sub-mm size of the filament also appears to be consistent with that observed by the microscopy for Pr$_{0.7}$Ca$_{0.3}$MnO$_{3}$ crystal.\cite{Fiebig1998ScienceA}
%
\par

\begin{figure}
\includegraphics[width=7cm]{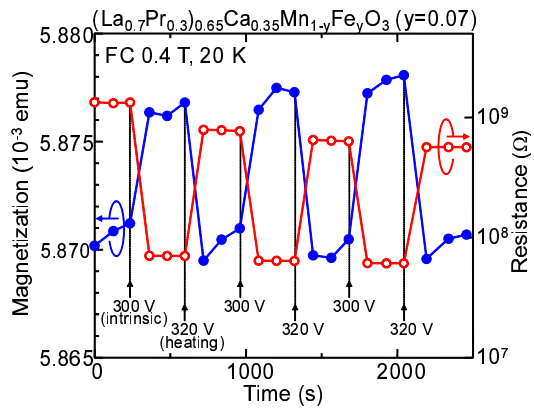}
\caption{\label{fig:switching}(Color online) Switching operation of magnetization (filled circle) and resistance (open circle) for a (La$_{0.7}$Pr$_{0.3}$)$_{0.65}$Ca$_{0.35}$Mn$_{0.93}$Fe$_{0.07}$O$_3$ single crystal ($d\!\sim\!50$ $\mu$m, $A\!\sim\!300\!\times\!400$ $\mu$m$^{2}$) at 20 K at 0.4 T after FC. The magnetization and resistance ($V_{\rm meas}\!=\!150$ V) were measured at regular intervals. Vertical arrows indicate the timing for applying the switching voltages.}
\end{figure}
%
Making use of the opposite actions of the intrinsic and extrinsic (heating) current-excitation effects, we demonstrate the cyclic electric control of both $M$ and $R$.
Figure \ref{fig:switching} shows the typical switching characteristics with changing the magnitude of the applied voltage, measured at 20 K at 0.4 T.
The detailed operation is as follows.
After FC to 20 K, we first checked that the application of $V_{\rm max}\!\le\!300$ V exhibited the intrinsic effect while $V_{\rm max}\!\ge\!320$ V was enough to cause the heating effect.
We then set the state by applying 320 V as the initial one, where the FM fraction was reduced by the aforementioned heating and subsequent quenching processes.
By applying 300 V to this state, a fraction of the FM state was recovered due to the intrinsic current-excitation effect, exhibited as an increase in $M$ (by $\sim\!6\!\times\!10^{-5}$ emu) and a decrease in $R$ (by $\sim\!1$ order of magnitude).
The subsequent application of 320 V gave again transient heating and decreased the FM component, returning $M$ and $R$ back almost to the starting values.
Repeating these two processes (``intrinsic" and ``heating") leads to the cyclic switching as shown in Fig. \ref{fig:switching}.
%
\par
%
This switching operation can be performed by using the voltage pulses (a few hundreds milliseconds), which is reminiscent of the colossal electroresistance memory (CERM) effect, as has been reported for the hole-doped CO manganites.\cite{Liu2000APLa,Sawa2004APLa}
The CERM phenomenon is the reversible resistance switching with bipolar (sometimes nonpolar) electric pulses, observed at room temperature for various transition-metal oxide films.\cite{Beck2000APLa,Seo2004APLa}
Its mechanism is still under controversy but one of the plausible scenarios is that a filamentary conducting path generated by the electric stress, such as the soft breakdown, is affected by the redox reaction due to the current heating effect.\cite{Ogimoto2007APLa}
Our result, on the other hand, presents the similar resistance switching by controlling the volume fraction of the (filamentary) metallic state, which can be regarded as a purely electronic CERM without suffering from the redox reaction (e.g. oxygen ion drift), although it works only at low temperatures at the moment.
%
\par
%
In conclusion, we have observed a persistent increase in a ferromagnetic-metallic (FM) volume fraction in a single crystal of (La$_{0.7}$Pr$_{0.3}$)$_{0.65}$Ca$_{0.35}$Mn$_{0.93}$Fe$_{0.07}$O$_3$ by applying voltages (300-500 V) between the electrodes (50-70 $\mu$m) .
The current-induced FM state is anticipated to form a filamentary pathway, where the net increase in magnetization is estimated as $\sim$0.4 $\mu_{\rm B}$ per Mn.
Utilizing the intrinsic current-excitation effect and the Joule heating one in a magnetic field, the latter of which conversely decreases the FM fraction, we have demonstrated reproducible switching of both the magnetization and resistance by changing the magnitude of the pulse voltage.
Such an electric control of the conducting and magnetic states for a relaxorlike manganite may find a route to oxide electronic devices in the future.
%
\begin{acknowledgments}
We thank R. Kumai, Y. Onose, and S. Iguchi for fruitful discussions.
This work was partly supported by MEXT TOKUTEI (16076205) and JSPS Fellows
\end{acknowledgments}
%

%
\end{document}